\documentclass[universe,review,accept,pdftex,moreauthors]{Definitions/mdpi} 

\usepackage{wrapfig, blindtext}
\usepackage{epsfig}
\firstpage{1} 
\pubvolume{12}
\issuenum{11}
\articlenumber{329}
\pubyear{2026}
\copyrightyear{2026}
\externaleditor{Firstname Lastname} % More than 1 editor, please add `` and '' before the last editor name
\datereceived{16 May 2026} 
\daterevised{22 October 2026} % Comment out if no revised date
\dateaccepted{22 October 2026 } 
\datepublished{30 October 2026} 
\Title{What the X-ray Emission from Young Supernovae Can Tell Us About Their Progenitors}
\Author{Vikram Dwarkadas \orcidA{} }

\AuthorNames{Vikram Dwarkadas}

\address[1]{%
 Department of Astronomy and Astrophysics, University of Chicago, Chicago, IL 60637, USA; vikramvd@uchicago.edu; Tel.:  +1-773-834-4668}
\abstract{We discuss constraints that the X-ray emission from supernovae can place on their progenitor stars. Since almost all X-ray supernovae arise from massive stars, we concentrate on implications for massive stars. These stars lose mass throughout their lifetime in the form of stellar winds. X-ray emission can be used to gauge the wind properties, especially the mass-loss rates, via various techniques. The abundance of various elements formed by nucleosynthesis within the stars, and released in the stellar explosion, can be estimated from X-ray spectra. In many cases, a combination of X-ray light curves and spectra can allow us to extricate the structure and morphology of the surrounding medium formed by pre-supernova mass loss into which the SN shock is expanding.}

\keyword{supernovae; supernovae: core-collapse; supernova: X-ray emission; X-Rays; massive stars;  stellar winds; mass-loss rates} 

\def\gtrsim{\mathrel{\hbox{\rlap{\hbox{\lower4pt\hbox{$\sim$}}}\hbox{$>$}}}}

\newcommand\msun{\rm M_{\odot}}

\newcommand\mdot{ \dot{M}}

\newcommand\be {\begin{equation}}
\newcommand\en{\end{equation}}

\newcommand{\ee}{\end{equation}}

\newcommand{\apj}{\mbox{ApJ}}
\newcommand{\apjl}{\mbox{ApJL}}

\newcommand{\araa}{\mbox{ARAA}}

\newcommand{\aap}{\mbox{AA}}
\newcommand{\gca}{\mbox{GeCoA}}

\newcommand{\mnras}{\mbox{MNRAS}}

\newcommand{\aapr}{\mbox{AApR}}

\begin{document}

%%%%%%%%%%%%%%%%%%%%%%%%%%%%%%%%%%%%%%%%%%
%\endnote{This is an endnote.} % To use endnotes, please un-comment \printendnotes below (before References). Only journal Laws uses \footnote.

% The order of the section titles is different for some journals. Please refer to the "Instructions for Authors” on the journal homepage.

\section{Introduction}

Core-collapse SNe (CCSNe) arise from massive stars, which lose mass throughout their lifetime in the form of winds. The wind is characterized by two main quantities, the mass-loss rate ($\dot{M}$) and the wind velocity ($v_w$). The wind  density ${\rho}_w \propto \dot{M}/v_w$. When the SN explodes, the SN shock expands within this wind-blown circumstellar medium, and not the pristine interstellar medium. The early emission from the SN is dependent on the evolution within this medium. One of the signposts of this circumstellar interaction is X-ray emission.  Almost all detected X-ray SNe, except for two, are of the core-collapse variety and, therefore, arise from massive stars \cite{dg12, dwarkadas25}.  For a general review of the X-ray emission from young SNe, we refer the reader to this review \cite{dwarkadas25}. A summary of the thermal and non-thermal emission from SNe is presented in reference \cite{cf17}.

In this paper, we will discuss the many ways that the X-ray data can help to constrain the SN progenitor. Since, as mentioned, almost all progenitors are massive stars, we concentrate on massive star progenitors and on SNe evolving in the medium created by mass loss from the massive star. We will discuss how to constrain the wind mass-loss parameters, decipher the structure of the surrounding medium, and evaluate the abundance distribution of the progenitor star, which can then be compared with various stellar explosion models. If the progenitor is part of a binary system,  the ambient medium may be a structured medium created during the binary evolution. Type Ia SNe are assumed to arise from thermonuclear explosion of a white dwarf in a binary system \cite{rs25}. If X-ray emission were to be detected from a garden-variety Type Ia-SN, similar constraints would be expected to apply to the companion stars. These would presumably be main-sequence or post-main-sequence low-mass stars ($< 8 \msun$),  giving rise to a higher-density medium surrounding the binary system, leading to detectable X-ray emission.  

The X-ray emission from young SNe can be either thermal or non-thermal. Thermal emission is due to thermal bremsstrahlung, which depends on the square of the density that the SN is expanding in. Non-thermal emission may be due to inverse Compton or synchrotron processes. At low ambient densities, the emission will be predominantly non-thermal. As the density of the surrounding medium increases,  thermal emission begins to dominate.  As shown by \cite{ci12}, inverse Compton cooling is also important at very high shock velocities and densities, although this has not been clearly observed. A review of the emission from various SN types is given in \cite{dwarkadas25}. Many SNe types show thermal emission. Some Type IIP SNe, which arise from red supergiant (RSG) stars, may evolve in a low-density medium (mass-loss rates $\dot{M} \approx 10^{-8} - 10^{-6} \, \msun$ yr$^{-1}$), in which case the emission is generally non-thermal. Type IIP SNe evolving in winds with higher ambient densities (larger $\dot{M}$) will show predominantly thermal X-ray emission. Non-thermal emission is also seen in Type Ib/c SNe, which are proposed to arise from Wolf-Rayet stars. These stars have wind velocities of 1000--3000 km s$^{-1}$. Since the wind  density ${\rho}_w \propto \dot{M}/v_w$, the wind density is low due to the high velocity, leading to non-thermal emission. Non-thermal emission is also expected from Type Ia SNe, which are not expected to evolve in a high-density medium. X-ray emission has not been detected from the standard garden-variety Type Ia SN.  The only Type Ia seen in X-rays, SN 2012ca \citep{bds18}, is classified as a Type Ia-CSM SN, which is one that shows an H line in its optical spectrum. The H is presumably due to mass loss from the companion star. The density of the X-ray emitting medium was found to be $> 10^6$ cm$^{-3}$, and perhaps as high as 10$^8$ cm$^{-3}$, and the emission is thermal. As this manuscript was being revised, another Type Ia-CSM SN, SN 2026sqf, has now been detected in X-rays \cite{marguttietal26}. Investigations of this SN are currently underway.

If the X-ray emission is thermal, it is directly proportional to the square of the circumstellar density \cite{cf17}. If the shock is radiative, as happens in many cases, then the emission is proportional to a single power of density. In either case, analysis of the X-ray emission can provide a direct handle on the density of the ambient medium. Evaluation of the X-ray luminosity over several epochs can help to determine the density profile of the medium in which the SN is expanding. In a CCSN, evaluation of the ambient density allows for a determination of the wind parameters. 

If the emission is non-thermal, such as synchrotron or inverse Compton, then the relationship with the external density is not direct. However, such a relationship can be established by connecting the relativistic particle energy density, or the magnetic energy density, to the thermal energy density. In the simplest case, this can be conducted by assuming equilibration between the different energy densities. More realistically, the relativistic particle and magnetic energy densities are assumed proportional to the thermal energy density, with the constant of proportionality being determined by other physics \cite{cf17}. The thermal energy density can be expressed in terms of the particle density, thus providing a handle on the external medium density.

Thus, one of the most significant constraints that the X-ray emission from a SN can provide is that it can tell us about the parameters of the progenitor star wind, which can then be subsequently related to the stellar parameters (Section \ref{sec:windpar}). We discuss various ways of doing this accurately in Section \ref{sec:diag}.

Thermal X-ray spectra often show lines of various elements. Fitting the spectra can enable us to evaluate the abundance of the specific element with respect to the solar value, as described in Section \ref{sec:abund}.  These abundances can also shed light on the composition of the progenitor star, thus allowing us to trace the progenitor.

The X-ray light curve and/or spectra can be used to determine the density profile of the ambient medium, and thus allow us to trace the structure and evolution of the medium. Asymmetries in the X-ray lines, or a combination of X-ray data with data at other wavelengths, can shed light on asymmetries in the surrounding medium, or the presence of a multi-component medium, as shown in Section \ref{sec:windbub}. The use of multi-wavelength data to evaluate wind parameters, especially the wind density, is discussed in Section \ref{sec:multiwv}. The density can then be tied to the wind parameters (Section \ref{sec:windpar}).  Final summary and discussion are covered in Section \ref{sec:disc}.

\section{Determination of the Progenitor Star Mass-Loss Rate}
\label{sec:diag}

\subsection{The Evolution of the CSM Density Profile and Thermal X-ray Luminosity}
In core-collapse SNe, the SN shock is initially expanding in the stellar wind formed by progenitor mass-loss \cite{chevalier82a}, characterized by the mass-loss rate $\dot{M}$ and wind velocity $v_w$.  For a constant mass-loss rate and wind velocity, the wind density decreases with radius as r$^{-2}$. Since thermal X-ray emission is directly dependent on the ambient density, the X-ray luminosity of core-collapse SNe also decreases with time if it is expanding in a constant parameter wind. The exception is when the expanding shock wave subsequently interacts with a dense medium such as a disk or wind-blown shell at later times, as in SNe 1987A \cite{cd95}, 1996cr \cite{baueretal08,ddb10}, SN 2004dk \cite{pooleyetal19} and 2014c \cite{thomasetal22,brethaueretal22}, leading to an increase in X-ray luminosity some time after the explosion. 

The evolution of the X-ray luminosity allows us to evaluate the density profile of the CSM. Since this density profile is formed by the mass loss from the pre-SN star, this can correspondingly inform us of the evolution of the pre-SN mass-loss rate. We use the \cite{chevalier82a} model to describe the evolution of the SN shock wave in the surrounding medium. This model assumes that the ejecta density profile can be described by a function that is a steep power-law $v^{-n}$ above a transition velocity $v_t$. Below this velocity, the density profile is relatively flat. The SN evolves homologously, and the ejecta density decreases with time t as t$^{-3}$. The value of $v_t$ is determined by integrating over the density profile to give the total ejecta mass, and preserving a finite kinetic energy. The SN ejecta density profile can be written as ${\rho}_{SN} = {\rho}_0\, {(v/v_0)}^{-n} \, {(t/t_0)}^{-3}$, and the uniform circumstellar medium into which the SN evolves has a density profile that decreases as ${\rho}_{CSM} \propto r^{-s}$ (s=2 for a steady wind with constant mass-loss parameters, s=0 for a constant density medium). The contact discontinuity then evolves as $R_{CD} \propto t^m$, where $m = (n-3)/(n-s)  \le$ 1. The thermal X-ray luminosity arising from a thin shell of thickness $\Delta R$ and radius $R$ between the forward and reverse shocks, both of which are assumed to be nonradiative,  can be shown to decrease approximately as \cite{flc96,dg12}

\begin{equation}
L_x \propto t^{-(12-7s+2ns-3n)/(n-s)} 
\label{eq:lcfull}
\end{equation}

The expression in Equation (\ref{eq:lcfull}) is valid when considering the total X-ray emission from the SN.  Most modern telescopes such as Chandra, XMM and Swift only measure the X-ray flux between about 0.5 -10 keV at best. If the shock temperature lies outside this region, or if it is high enough that emission falls outside the region, measurements with Chandra, XMM, or Swift will not provide an accurate total X-ray luminosity. X-ray temperatures may exceed 10 keV at early times. SN 1993J was observed with the OSSE telescope 12 days after explosion with a luminosity of 5 $\times 10^{40}$ erg s$^{-1}$ in the 50--150 keV energy range \cite{leisingetal94}. SN 2023ixf had temperatures $> $ 30 keV during the first month \cite{nayanaetal25}. SN 2014c also had temperatures exceeding 10 keV \cite{thomasetal22,brethaueretal22}. In the latter two cases, the total X-ray luminosity was obtained with the help of the NuStar telescope, which measures the emission from 3 to 79 keV and can capture the emission from SNe where the temperature exceeds 10 keV. If the temperature exceeds 10 keV, but the luminosity beyond 10 keV has not been measured, then Equation~(\ref{eq:lcfull}) would not be valid and would result in an inaccurate determination of $s$. It should be noted that the NuStar telescope has not detected many X-ray SNe.

For a steady wind ($s=2$), Equation (\ref{eq:lcfull}) reduces to $L_x \propto t^{-1}$. Thus, for a steady wind, the X-ray emission is  independent of the ejecta density profile and decreases inversely with time. We caution that this is valid only if the entirety of the X-ray emission is captured, which as mentioned above is not often true for young SNe.

If the emission lies in an energy band $E << kT_{sh}$, where $T_{sh}$ is the shock temperature, then we can use Equation (\ref{eq:lcband}) instead to describe the optically thin X-ray luminosity due to thermal bremsstrahlung  in terms of the ejecta and surrounding medium density \cite{flc96}. 

\begin{equation}
L_x \propto t^{-(6-5s+2ns-3n)/(n-s)} \;\;\;\;\;  E << kT_{sh}
\label{eq:lcband}
\end{equation}

The value of $n$ was investigated by \cite{matzner99} and found to lie between $9 \leq n \leq 12$ for a large number of SN progenitors, assuming spherical symmetry. The X-ray luminosity $L_x$ for a steady wind in this case will evolve as t$^{-0.7}$ to t$^{-0.8}$.  For larger values of the ejecta slope $n$, the luminosity evolution becomes progressively closer to $t^{-1}$. This is the lightcurve slope that would be expected for SNe with a high temperature $ > 10$ keV, if the X-ray emission was measured with Chandra, XMM-Newton or Swift. Since the post-shock temperature is a function of the shock velocity for a strong shock, high temperatures are more likely during the early evolution when the SN shock velocities are high.

Having calculated the X-ray luminosity $L_x$ with time (the X-ray light curve), the value of `s' can be derived from Equations (\ref{eq:lcfull}) or (\ref{eq:lcband}). The density of the ambient medium, for a mass-loss rate that is constant or gradually varying with time, can be written as 

\begin{equation}
{\rho}_{am} = \frac{\dot{M}}{4 \pi {r_o}^2v_w}\left[\frac{r_o}{r}\right]^{s}
\end{equation}

\noindent
where $r_0$ is a fiducial radius at which the wind parameters are computed. If the mass-loss parameters are constant, $s=2$, and the $r_0$ term drops out, since the wind parameters are the same at all radii. Assuming a value of the wind velocity $v_w$ depending on the type of SN, and knowing the shock radius (inferred by other means), the value of the mass-loss rate $\dot{M}$ can be obtained.  The velocity $v_w$ is of order 10 km s$^{-1}$ for red supergiant progenitors, 100 km s$^{-1}$ for the progenitors of Type IIn SNe, and  1000--3000 km s$^{-1}$ for the Wolf-Rayet star progenitors of Type Ib/c SNe \cite{marcowithetal18}. Thus, the evolution of the mass-loss rate at different epochs can be obtained, which can be related to the properties of the progenitor star.  The mass-loss may also be non-uniform or explosive in nature, as considered in later sections.

\subsection{Mass Loss Rates Correct to Order of Magnitude}

\textls[-15]{If the X-ray emission from a SN  is thermal, the luminosity of the thermal X-ray emission from a non-radiative SN shock evolving in a steady wind can be written as (\cite{CF03}):}

\begin{equation}
L_i = 3 \times 10^{39} g_{ff} C_n \left(\frac{{\mdot}_{-5}}{{v_w}_{1}} \right)^2 \left(\frac{t}{10 \,{\rm d}} \right)^{-1} {\rm ergs\; \;s^{-1}}
\label{eq:xraylum}
\end{equation}

\noindent
The equation is applicable to both forward and reverse shocks, depending on the value of $C_n$.  $C_n = 1$ for the circumstellar (forward) shock,  and $C_n=(n-3)(n-4)^2/4(n-2)$ for the reverse shock. $g_{ff}$ is the Gaunt factor. ${\mdot}_{-5}$ is the mass loss rate of the wind into which the SN is expanding, in units of 10$^{-5}$ $\msun$ yr$^{-1}$; ${v_w}_{1}$ is the wind velocity in units of 10 km s$^{-1}$ (characteristic of RSGs), and $t$ is the time in days since the SN explosion. The mass loss rate and wind velocity are assumed constant, leading to a wind density decreasing as r$^{-2}$, and an X-ray luminosity decreasing as $t^{-1}$, as described above. We again emphasize that the X-ray luminosity refers to the total luminosity of the X-ray emission, while current X-ray satellites except for NuStar are mainly effective in a limited X-ray range between 0.5 and 10 keV. Electron-ion equilibration behind the shock front is assumed. However, the equation can be modified to account for any ratio of electron to ion temperature ($T_e/T_p$).

\begin{figure}[H]
\includegraphics[width=0.9\columnwidth]{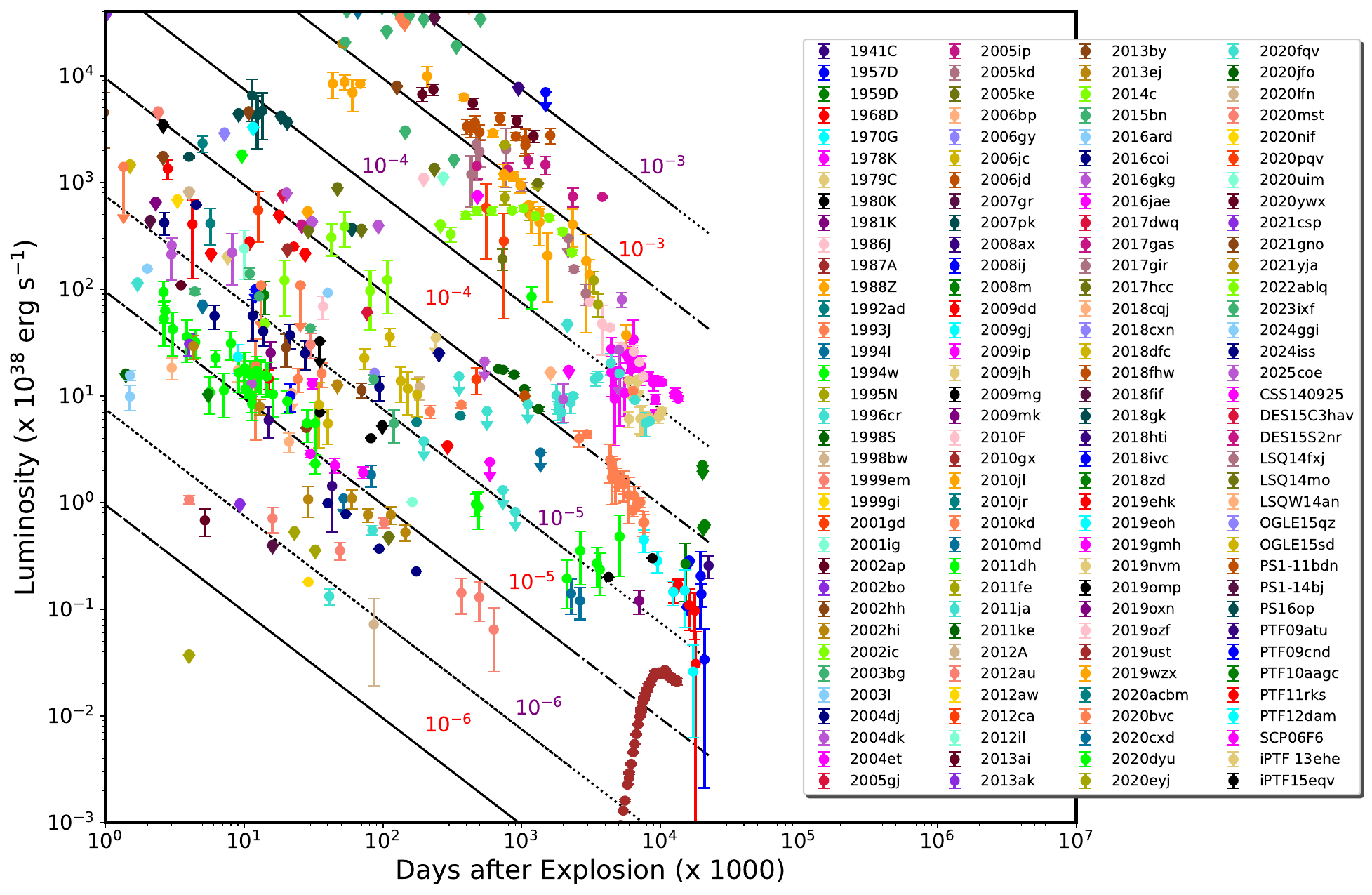} 
\caption{X-Ray Lightcurves of most SNe observed at X-ray wavelengths. There are 140 SNe shown here. Solid and dashed lines  represent constant mass loss, with  t$^{-1}$ slope (Equation (\ref{eq:xraylum}) \cite{CF03}). Mass loss rates in $\msun$ yr$^{-1}$. Mass loss rates in red assume that the emission arises from the forward-shocked steady wind  (density $\propto$ r$^{-2}$), v$_{\rm wind}$ = 10 km s$^{-1}$, electron temperature 10\% of post-shock ion temperature. Mass loss rates in purple make similar assumptions, but the X-ray emission is assumed to arise from reverse-shocked ejecta, which as shown in Equation (\ref{eq:xraylum}) will result in a higher level of emission than the forward shock. A value of $n=10$ for the ejecta density profile is assumed. 
\label{fig:lcmdot}}
\end{figure}

Figure \ref{fig:lcmdot} shows the X-ray light curves of most X-ray SNe, excluding some superluminous SNe, which have mainly upper limits. This figure is adapted from a similar figure in \cite{dwarkadas25}, with more SNe added. Inverting Equation (\ref{eq:xraylum}), lines of X-ray luminosity at a  constant mass loss rate are plotted. $T_e/T_p$ = 0.1 is used, representative of many young SNe.  Solid lines in the figure denote a fixed mass-loss rate, listed in red, assuming the X-ray emission arises from the forward shock ($C_n = 1$). 

 In young SNe, the forward shock is generally expanding at several thousand km s$^{-1}$. A shock expanding at 5000 km s$^{-1}$ would result in a post-shock temperature of about 40 keV. Faster shocks would result in even higher temperatures. However, fits to the X-ray spectra show that the X-ray temperature in many SNe is found to be on the order of 1--2 keV, much lower than the post-shock temperature of the forward shock. Although electron-ion temperature inequilibrium can reduce the observed electron temperature, the low electron temperature suggests that the emission arises primarily from the reverse shock \cite{CF03,chandraetal09,maedaetal14,gd26}. The reverse shock is expanding into much higher density, thus its velocity, and therefore, the temperature behind the shock is lower. Figure \ref{fig:lcmdot} also includes lines of constant mass-loss rate denoted in purple, when the X-ray emission is assumed to arise from the reverse shock, rather than the forward shock. A value of $n=10$ is used. Note that the value of $C_n$ for the reverse shock is always greater than 1, and thus the X-ray luminosity from reverse-shocked ejecta, for the same mass-loss rate, is higher than that from forward shocked circumstellar medium. This is due to the fact that the reverse shock is moving up a steep density incline, at least in the early phases, and is interacting with a much higher density than the forward shock. The larger the value of $n$, the higher the X-ray luminosity for the same mass-loss rate. Thus, it is possible that two SNe expanding in a medium with the same mass-loss rate could have different X-ray luminosities if the emission arises from reverse-shocked ejecta with different values of the ejecta density parameter $n$.

As seen from Equation (\ref{eq:xraylum}), what is actually obtained in practice is not the mass-loss rate itself but the parameter $\dot{M}/v_w$ (Equation~(\ref{eq:xraylum})). Rates are quoted for $v_w = 10$ km s$^{-1}$. For an arbitrary wind velocity $v_w$, the mass-loss rate should be multiplied by  $v_w/10$. 

For SNe evolving in a steady wind, whose luminosity closely follows the $t^{-1}$ decrease, mass-loss rates deduced via this method are close to those obtained by other means. Figure \ref{fig:lcmdot} suggests a mass-loss rate  $ > 10^{-5}\; \msun$ yr$^{-1}$ for SN 1993J in the early stages, close to the rate of 4 $\times 10^{-5} \;\msun$ yr$^{-1}$ deduced by \cite{VT09} via radio lightcurve fitting.

\textbf{Implications for the Progenitors}: The plot reveals that the X-ray luminosity of many SNe does not decrease with time as t$^{-1}$, or conversely that they do not evolve in a  steady wind, but one whose parameters change with time. This is especially true of Type IIn SNe, which show a steep decrease in X-ray luminosity after a few years, as had already been noted by reference \cite{dg12}.  Assuming a constant wind velocity, this implies that the mass-loss rate is increasing as the star moves closer to explosion. 
In some SNe an increase in X-ray luminosity is observed some time after the explosion, e.g., in SN 1987A \cite{cd95}, SN 996cr \cite{baueretal08,ddb10}, SN 2004dk \cite{pooleyetal19} and SN 2014c \cite{thomasetal22,brethaueretal22}.  The progenitor mass-loss may have undergone a change in wind parameters, and the CSM density profile will be different from r$^{-2}$. It may resemble a wind-blown bubble, formed by a faster wind sweeping up a slower wind from a previous epoch. It could also consist of an equatorial disk or torus, or high-density clouds. When the shock impacts this high-density region, it gives rise to increasing X-ray emission. These are discussed further in Section~\ref{sec:windbub}.

\subsection{Accurate Determination of Mass Loss Rates}
\label{sec:detml}

A more accurate mass loss rate determination from the X-ray emission, correct to within a factor of 2,  requires the SN spectrum rather than just the light curve.  This method is particularly relevant if the SN is not evolving in a steady wind with  constant properties. The mass loss properties are then a function of radius, or equivalently time after explosion.  The time evolution of the luminosity can be related to the density structure of the surrounding medium, as shown in \cite{flc96} and
\cite{dg12}. The \cite{chevalier82a}
description for a SN shock wave evolving in a self-similar spherically symmetric manner is used here. The procedure outlined in \cite{flc96} is followed to calculate the mass loss rate. Since the rate would be a function of radius (or time),  the quantity $\dot{M}/v_w$ is computed at a specific reference radius, assumed to be 10$^{15}$ cm, from which it can be derived at any other radius.  Following \cite{flc96} and \cite{drrb16}, a spectral
luminosity at 1 keV is used.  A value of 1 keV is chosen as there are no lines present at that precise energy value.  The forward shock luminosity at 1 keV can be written as follows:

\begin{equation}
%\begin{split}
L_{cs, 1keV} = 1.4 \times 10^{38} \,\xi \,T_8^{-0.236} \frac{e^{-0.116/T_8}}{(3-s)}\left[\frac{{\dot{M}_{-5}}}{v_{w1}}\right]^2  V_4^{3-2s}\,\left[\frac{t_d}{11.57}\right]^{3-2s} {\rm ergs \, s^{-1} \, keV^{-1}}\;\; .
%\end{split}
\label{eq:lcs}
\end{equation}

\noindent
where $V_4$ is the maximum ejecta velocity scaled to 10$^4$ km s$^{-1}$,
$\xi = [1 + 2n(He)/n(H)]/[1 + 4 n(He)/n(H)] \sim 0.85$, and $t_d$ is
the time in days. This equation is valid for an arbitrary value of $s$. Unlike Equation~(\ref{eq:xraylum}), it is not independent of the value of $n$, although  the variable $n$, or the ejecta density constant ${\rho}_0$, does not appear explicitly. Instead, the maximum ejecta velocity $V$, which is a function of $n$, is preferably used, since it serves as a better observable. A similar expression can be derived for the reverse shock.

{\bf Examples:} With well-sampled X-ray light curves and moderate-resolution spectra, the mass loss rates of individual SNe can be determined very accurately \cite{flc96}. \cite{chandraetal15} determined a mass loss rate of 0.06 $\msun$ yr$^{-1} $ for SN 2010jl, in agreement with the optical value of 0.1 $\msun$ yr$^{-1}$ \cite{franssonetal14}. 

\begin{figure}[H]
\includegraphics[width=\columnwidth]{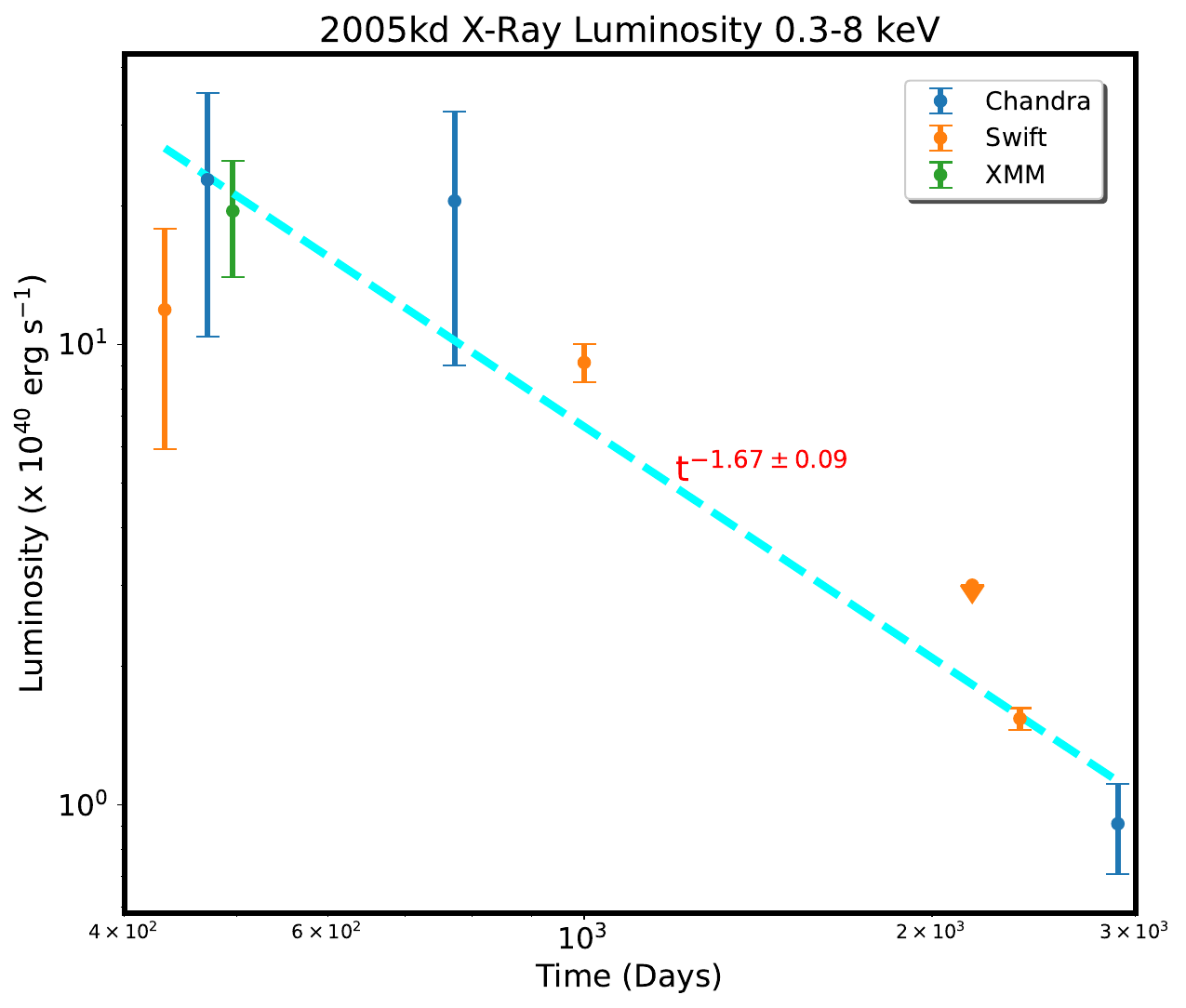}
\caption{The X-ray lightcurve of SN 2005kd, using the luminosity values from \cite{drrb16}. The luminosity is found to decrease with time $t$ as t$^{-1.67 \pm 0.09}$. Note that this value of the luminosity decrease is somewhat different from that of \cite{drrb16}, presumably due to differences in the fitting algorithm, although it is consistent within the error bars. Using their value for the decrease, the X-ray luminosity at 1 keV from the spectrum, and Equation (\ref{eq:lcs}), \cite{drrb16} found an accurate mass loss rate as a function of time, or equivalently distance from the SN explosion.
\label{fig:05kdmdot}}
\end{figure}

For SN 2005kd, \cite{drrb16} find a mass loss rate of 4.3 $\times 10^{-4} \;\msun$ yr$^{-1}$ with $v_w$= 10 km s$^{-1}$ at 10$^{16}$ cm (see Figure \ref{fig:05kdmdot}). Closer to core-collapse it could have been $> 10^{-3}\; \msun$ yr$^{-1}$. 

\subsection{Mass-Loss Rate from Column Density Variations:} For a SN evolving in a wind medium with density decreasing as r$^{-s}$, the column density of the medium around the SN can be written as \cite{drrb16}:

\be
N(H)_{cs} = 2.1 \times 10^{22} \left[\frac{\dot{M}_{-5}}{(s-1)\,v_{w1}}\right]\,V_4^{1-s}\left[\frac{t_d}{8.9}\right]^{1-s} {\rm cm}^{-2} \;\;.
\label{eq:column}
\ee

This is the additional column that must be added to the value of the Galactic column density. Even if there is some other component that contributes to the column density, such as the column through the host galaxy, or between the host galaxy and the observer, the change in the column density between two observations is likely due to the shock having advanced through the surrounding medium, and can be used to determine the mass-loss rate.

\subsection{Inverse Compton Emission:} 
\label{sec:invc}
The emission from the SN need not be thermal, as mentioned above, but could be non-thermal. The non-thermal emission may have either a synchrotron or inverse Compton origin. \cite{cfn06} suggested that Type Ib/c SNe may show inverse Compton X-ray emission early on. \cite{cf06} suggested that the X-ray emission from low X-ray luminosity Type IIP SNe may also be due to inverse Compton. Assuming that
the spectrum of accelerated electrons follows a power law N(E) $\propto E^{-p}$,
we can write for the case p=3, and for a SN expanding into a steady
wind:

\noindent
\be
E \frac{dL_{IC}}{dE} \approx 8.8 \times 10^{36} {\gamma}_{min}\,S_{\odot} {\alpha}^{11/19} v_{s4} \left[\frac{L_{bol}(t)}{10^{42}\; {\rm ergs}\; {\rm s}^{-1}}\right] t_{10}^{-1}\; {\rm ergs}\; {\rm s}^{-1}
\ee
 
\noindent
where $L_{IC}$ is the Inverse Compton Luminosity, $\gamma_{min}$ is
the Lorentz factor of the relativistic electrons, $ v_{s4}$ is the
forward shock velocity in units of 10$^4$ km s$^{-1}$, $L_{bol}$ is
the bolometric luminosity of the SN, ${\alpha}$ is the ratio of
relativistic electron energy density to magnetic energy density,
$S_{\odot} = {\epsilon}_{B-1} A_{\star} \alpha^{8/19}$,
${\epsilon}_{B-1}$ is the ratio of magnetic energy density to thermal
energy density, wind density $\rho = A r^{-2}$ and A$_{\star} = A/(5
\times 10^{11}) \,{\rm g}\, {\rm cm}^{-1}$. This expression can be
modified for an arbitrary value of $p$, and for evolution in arbitrary
media.

For a population of optical photons with effective temperature T$_{eff}$, the IC luminosity at a frequency $\nu$ reads \cite{marguttietal12}:

\noindent
\be
\label{eq:ic}
E \frac{dL_{IC}}{d\nu} \approx 0.2 \left(\frac{h}{3.6k}\right)^{(3-p)/2} \times \frac{(p-2){\sigma}_T  {\epsilon}_e    {\rho}_{CSM} v_s^2 {\gamma}_{min}^{p-2} {T_{eff}}^{(p-3)/2} {\nu}^{(1-p)/2} \Delta R}{m_e c^2} L_{bol}(t)
\ee
 \noindent
where $\Delta$R is the extension of the region containing fast electrons; ${\rho}_{CSM} \propto R^{-s}$ is the circumstellar medium density the SN shock is expanding into, parameterized as a power law in shock radius; the shock velocity v$_s$ = v$_s$(t, n, s); the shock radius R = R(t, n, s) and ${\gamma}_{min}$ = ${\gamma}_{min}$(t, n, s)

\textls[-15]{Knowing the IC luminosity, these equations can provide the CSM density. IC emission requires initial seed photons that, on impact with
accelerated electrons, are upscattered to higher energies. Although
\cite{cf06} assume some fraction of the bolometric luminosity, in
well-observed cases the energy density of the visible or infrared
components could be used instead.}

{\bf Examples:} Knowing the IC luminosity, and using  equations in \cite{cfn06, marguttietal12}, the CSM density and other parameters can be calculated. \cite{marguttietal12} have used the IC luminosity to constrain the mass loss rate of SN 2011fe, a Type Ia SN, using upper limits on the X-ray emission. For those SNe where X-ray emission is clearly detected, more accurate calculations of the mass loss rate can be made. For SN 2011ja, \cite{sayanetal13} deduced a mass loss rate of 1.2 $\pm \; 0.2 \times 10^{-7}\; \msun$ yr$^{-1}$. 

\subsection{Radiative Shocks} \textls[-25]{Radiative shocks may occur in young SNe, when the ambient medium density is very high and the cooling time is low. This can be determined by comparing the cooling time behind the shock to the flow time, or simply to the age of the remnant.
 It is possible for the reverse shock to be radiative whereas the forward shock is
not, as determined for SN 1993J \citep{nymarketal09, chandraetal09}. 
}
The cooling time, for a shock temperature $< 2 \times 10^7$, and a CSM density decreasing as r$^{-s}$, is \cite{flc96}:

\be
t_{cool, r} = 3.5 \times 10^9 \frac{(4-s)(3-s)^{4.34}}{(n-3)(n-4)(n-s)^{3.34}}\, 
V_4^{3.34 + s} \left[\frac{\dot{M}_{-5}}{v_{w1}}\right]^{-1}\left[\frac{t_d}{11.57}\right]^{s} s \;\;.
\label{eq:rcool}
\ee

For $kT > 2 \times 10^7~\mathrm{K}$ 

\begin{equation}
t_{\mathrm{cool}} = \ 1.6 \times 10^3 \, \frac{(4-s)(3-s)^{2}}{2(n-s)(n-3)(n-4)}  \times V_4^{\,5 - s}  \left( \frac{\dot{M}_{-5}}{v_{w1}} \right)^{-1} \left( \frac{t}{11.57~\mathrm{d}} \right)^{4-s} \ \mathrm{d}
\label{eq:coolhigh}
\end{equation}

\noindent
Comparing this time to the age of the remnant can tell whether the shock is cooling. 

\textls[-15]{The total luminosity of a radiative reverse shock is given by Equation 3.17 in \citet{flc96}}

\begin{equation}
\begin{aligned}
\label{eqn:rad}
    L_{rev} = 1.57 \times 10^{41} \,\frac{2(3-s)^2(n-3)(n-4)}{(4-s)(n-s)^3} \, \left(\frac{\dot M_{-5}}{v_{w1}}\right) \times V_4^{5-s} \, \left(\frac{t}{11.57 \; \text{d}}\right)^{2-s}
\end{aligned}
\end{equation}

\noindent
The luminosity of a nonradiative reverse or forward shock is given by Equation (\ref{eq:xraylum}).

\textls[-15]{Equations (\ref{eq:xraylum}) and (\ref{eq:rcool})--(\ref{eqn:rad}) all depend on the mass-loss rate. Thus, in order to figure out the correct mass-loss rate, the mass-loss rate must first be determined from both \mbox{Equations (\ref{eq:xraylum}) and (\ref{eqn:rad})}. This is then input into Equations (\ref{eq:rcool}) and (\ref{eq:coolhigh}) to calculate the cooling time. If the cooling time is found to be lower than the age of the SN, the mass-loss rate corresponding to the radiative case is used, the mass-loss rate corresponding to the adiabatic case is used.}

\textls[-15]{Using the above procedure for SN 2011dh, \cite{gd26} showed that the X-ray emission arose from an adiabatic reverse shock. They found the mass-loss rate to be around a few times 10$^{-6} \; \msun$ yr$^{-1}$.}
 
\subsection{Variable and Episodic Mass Loss}
\label{sec:varmass}

Variations in the X-ray light curve over time can provide information on density variations in the surrounding medium, which can arise from mass loss rate variations. (Note that not all density fluctuations necessarily indicate mass loss rate fluctuations, see Section \ref{sec:windbub}). Thus, analysis of the X-ray light curve can provide information on mass loss variations in the pre-SN star.

\begin{figure}[H]
\includegraphics[width=\columnwidth]{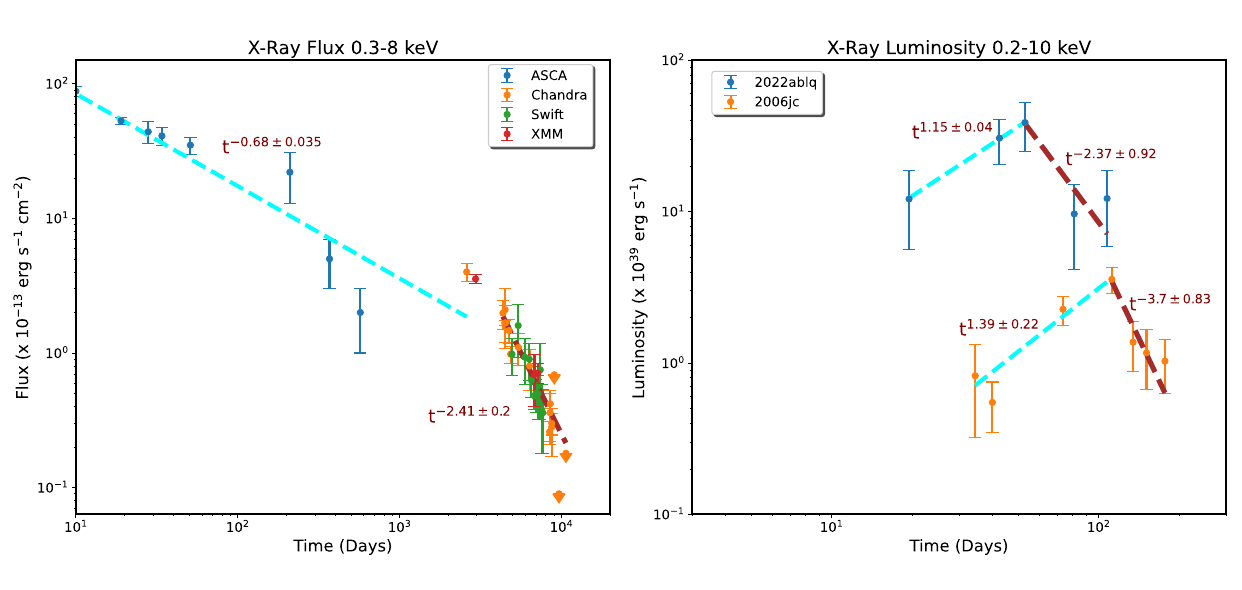}
\caption{[Left] The X-ray light curve of SN 1993J, which shows a break at around 3500 days. This plot was made by analyzing all available X-ray data on SN 1993J \cite{dwarkadas26}, with an increase in the time baseline of $\approx$ 10 years compared with the data plotted in \cite{dbbb14}. The drop in flux is likely due to a change in the mass-loss properties. \cite{dbbb14, kunduetal19}. The initial ASCA data points are not considered as reliable as they may include emission from a nearby source. [Right] The X-ray lightcurve of SN 2022ablq compared with that of SN 2006jc \cite{pellegrinoetal24}. Both show an increasing X-ray luminosity for $\approx$ 70 days, followed by a decreasing flux for the next couple of months.
\label{fig:varmdot}}
\end{figure}

\textbf {Examples:} The X-ray light curve of one of the best observed SNe, the Type IIb SN 1993J, shows that the X-ray flux decreased abruptly after about 3500 days, as first reported in  \cite{dbbb14}. At this time, the slope of the light curve changed dramatically (Figure \ref{fig:varmdot}, left). The radio light curve also shows a similar change in slope \cite{Weiler2007}. These appear to indicate that the mass loss rate changed dramatically \cite{dbbb14,kunduetal19}. The progenitor of SN 1993J is known to be a binary, and it is possible that the change in mass loss could be related to changes in the accretion rate or mass transfer to the binary progenitor.  

\noindent

Reference \cite{pellegrinoetal24} showed that the light curve of SN 2022ablq increased for more than 2 months, then suddenly began to decrease. This is similar to the well-known SN 2006jc. These changes are presumably due to changes in the wind properties. The type IIP SN 1999em also showed fluctuations in its light curve, which \cite{pooleyetal02} argued were due to variations in the mass loss rate. 

The wind variations mentioned here are sudden, drastic variations in the mass loss properties. This is to be contrasted with a gradual variation due to smooth time-dependent mass loss, leading to a density profile varying as r$^{-s}$, as discussed in Section \ref{sec:detml}.

\section{Ejecta Abundances}
\label{sec:abund}

The explosion of a  core-collapse star leads to the expulsion of the stellar envelope, containing the nucleosynthesis products formed during nuclear burning or during the explosion. This ejected material, or SN ejecta, can be considerably enriched in metals such as Si, Ca, and  Fe, depending on the mass of the star. The abundance is reflected in thermal X-ray spectra and can be measured in adequate-resolution spectra.  It can then be compared with that obtained from detailed SN explosion models, to provide clues to the progenitor. 

\textls[-15]{In a young SN, the reverse shock is gradually moving back into the ejecta (in a Lagrangian sense), and has not thermalized the entire ejecta. Only emission from the shocked ejecta material is generally seen. It is possible that X-ray emission from the reverse shock can ionize the unshocked ejecta, which can then recombine, giving rise to observable emission. In spherical symmetry, one expects the stellar envelope to be layered, with heavier elements towards the inside. As the reverse shock moves inwards, ejecta containing heavier metals will be sequentially shocked. Mixing of the ejecta due to turbulence and hydrodynamic instabilities introduces further complications to this picture. A series of spectra over several epochs can better constrain the ejecta, such as  in SN1996cr (see Table 4 in \cite{qbd19})}.

\begin{figure}[H]
\includegraphics[width=\columnwidth]{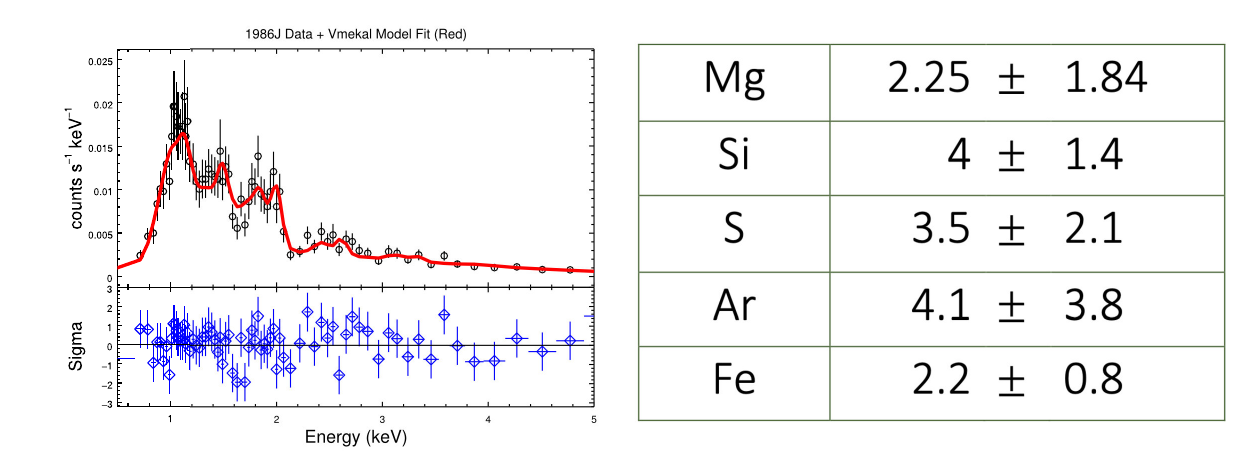}
\caption{ (Left) ACIS-S spectrum of SN 1986J, fitted with a two-component model. (Right) Best-fitting abundance distributions, scaled to solar abundances, from the fit.  
\label{fig:86J}}
\end{figure}

Determination of abundance is possible in very high-resolution transmission grating spectra, e.g., in SN 1987A \cite{zhekovetal06,zhekovetal09,deweyetal12} and SN 1996cr \cite{qbd19}. However, it is not restricted only to grating spectra. It is also achievable with high-resolution spectra using the Chandra ACIS-S instrument or the MOS1, MOS2 and PN telescopes on XMM-Newton. By high-resolution, we mean that enough counts are present to clearly delineate lines of various elements. Figure \ref{fig:86J} shows an archival ACIS spectrum of SN 1986J, along with a two-component fit, and the resulting abundance determination.

\noindent

In X-ray spectral fitting software, the abundance of an element is the number ratio of that element to that of H. Abundances are scaled to a fiducial solar standard, generally that defined by \citet{ag89}, although different standards can be used. The fits assume that the abundance of an element is equal to this standard solar value, unless the element is `thawed' and its abundance is allowed to vary. The values quoted in Figure~\ref{fig:86J} are the number ratio of the element abundance to the solar value.

\begin{figure}[H]
\includegraphics[width=\columnwidth]{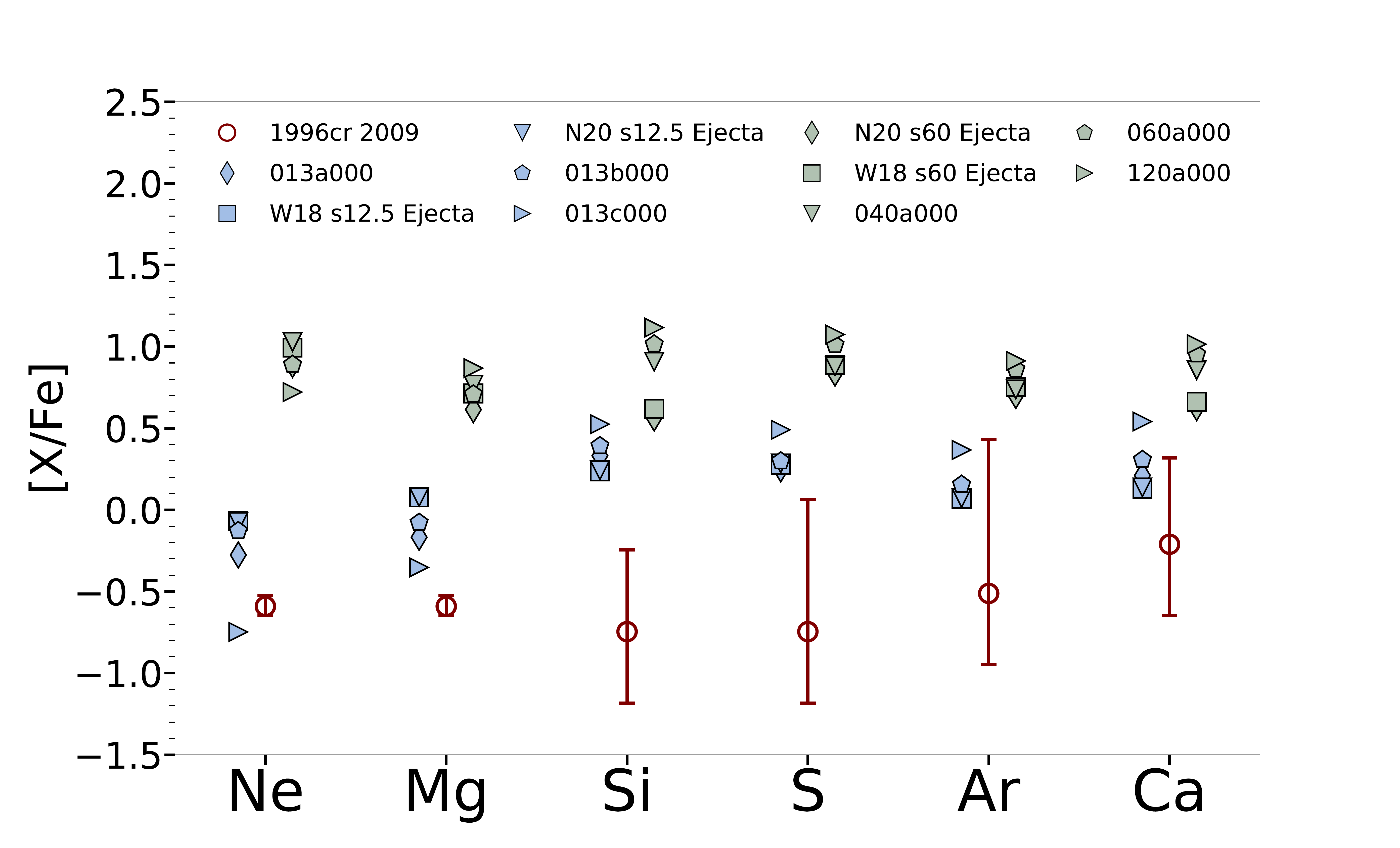}
\caption{Comparison of the 1996cr 2009 epoch reverse-shocked data (signifying ejecta) with a selection of core-collapse SN models.  Although none of the explosion models fit all the elements, low initial mass core-collapse models with masses between 12 and 15 $\msun$ appear to provide the best fit.
\label{fig:abundcomp}}
\end{figure}

\textls[-15] We can then compare the observed abundances to ejecta abundances computed from a large selection of SN explosion models published in the literature, to determine the mass of the star that fits best. Figure \ref{fig:abundcomp} shows the comparison of abundances obtained for 1996cr with a  variety of explosion models for core-collapse SNe available in the literature.  The various explosion models are further described in \cite{siegeletal20b}.  Low initial mass core-collapse models provide the best fit for 1996cr; higher mass models tend not to fit. Coupled with the fact that the progenitor was likely a W-R star or an 87A-like blue supergiant \cite{ddb10}, this appears to suggest evolution in a binary system. The presence of an asymmetry in the shapes of various lines (Figure 3 in \cite{qbd19}) in the grating spectra of SN 1996cr, suggesting an asymmetry in the ejecta or surrounding medium or both, also pointed to binary evolution \cite{qbd19}.

\noindent

Abundance comparisons such as this have been carried out for supernova remnants (SNRs) \cite{lopezetal13, zv18, siegeletal20a, siegeletal20b, shardaetal20}, but not as often for young SNe.

\section{Investigating the Structure and Morphology of the Ambient Medium}
\label{sec:windbub}
\begin{figure}[H]
\includegraphics[width=17 cm]{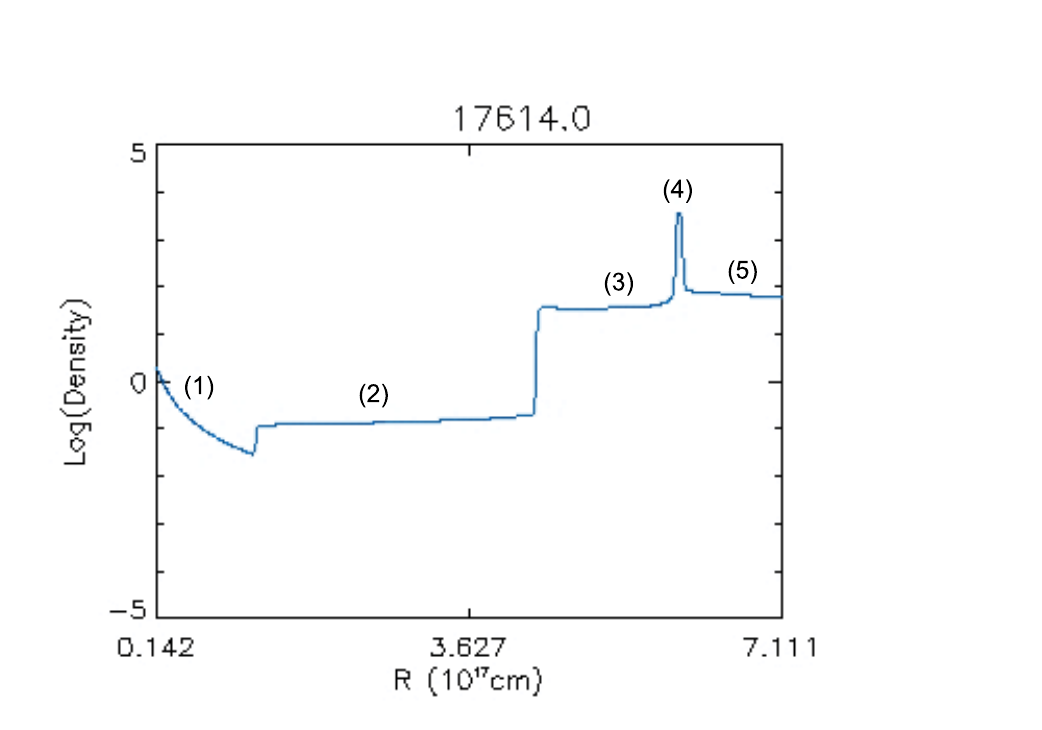}

\caption{The density structure of the medium around SN 1987A in the equatorial plane, using a simulation that combines the photo-ionizing photons from the central pre-SN star with the hydrodynamics of the wind bubble evolution. The inclusion of the stellar photons results in the formation of a dense photo-ionized HII region interior to the shell. Going outwards in radius from the central star, the various regions are (1) the freely expanding wind, (2) the shocked wind, (3) the photo-ionized (HII) region, (4) the dense swept-up shell (equatorial ring) and (5) the red-supergiant wind from the previous epoch. Details of the simulation and the modeling technique are described in \cite{vvd22}.
\label{fig:87afinal}}
\end{figure}

\begin{figure}[H]

\begin{adjustwidth}{-\extralength}{0cm}
\centering 
\includegraphics[width=18cm]{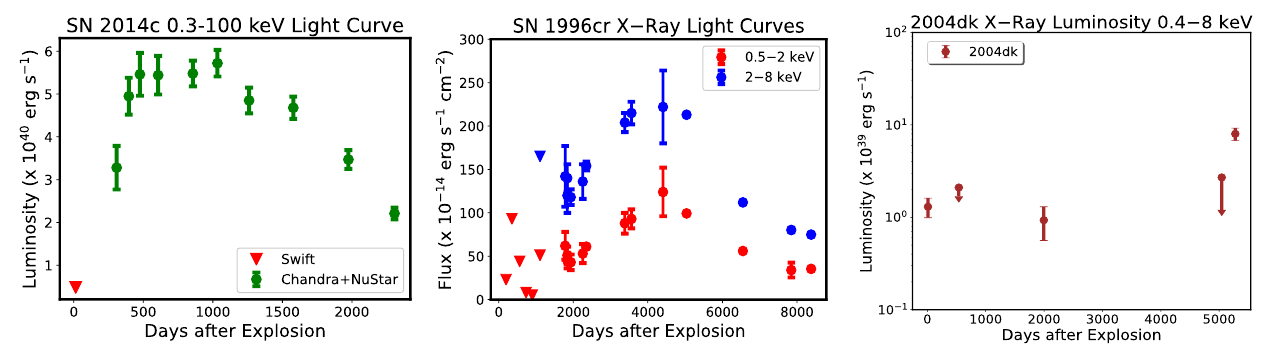}
\end{adjustwidth}
\caption{Examples of X-ray light curves of Type Ib/c SNe that show a rise in X-ray emission, indicating the presence of an increased density in the circumstellar medium. (\textbf{Left}) Type Ic SN 2014c. Data from \cite{thomasetal22}. (Middle) 0.5--2 and 2--8 keV lightcurves of Type IIn SN 1996cr (presumed to be a Type Ib/c at explosion). Adapted from \cite{qbd19}.  (\textbf{Right}) Type Ib SN 2004dk. Data from \cite{pooleyetal19}.
\label{fig:windbubxray}}
\end{figure}

The X-ray light curve can be used to investigate the structure and morphology of the surrounding medium, which was formed during the pre-SN evolution. Stellar wind properties can change dramatically in the years to centuries before the star undergoes core-collapse \cite{foley2007, smith17hsn}. Variations in the stellar wind properties can lead to a faster wind sweeping up the slower wind from a previous epoch, creating massive wind-blown cavities around a star, bordered by a dense shell \cite{weaver1977}. These wind-blown bubbles should technically form around most massive stars. Their size and the shell density vary considerably depending on the wind properties.  The SN shock will initially propagate into a low-density medium carved out by the stellar wind shocks, before eventually colliding with the dense shell of swept-up ambient material enclosing the bubble (see Figure~\ref{fig:87afinal}). The collision of the SN shock with the high-density shell will lead to a sharp increase in the X-ray emission (Figure~\ref{fig:windbubxray} \cite{Dwarkadas2005, Dwarkadas2007c}). The intensity of the X-ray emission will vary depending on the density contrast between the interior of the wind bubble and the dense shell.  Analysis of wind bubbles can provide estimates of not only the progenitor mass loss, but also the mass loss rate of the wind in the prior epoch \cite{ddb10}, into which the progenitor wind was expanding \cite{ddb10, pooleyetal19}.  

\vspace{-12pt}
\noindent

\noindent

An increase in X-ray emission will also happen if the SN shock collides with a high-density medium, such as a dense disk or torus in the equatorial plane, a dense HII region and equatorial ring as in the case of SN 1987A, or high-density clouds. A model to explain the emission from Type IIn SNe as a result of a collision with a dense disk or high-density clouds embedded in the surrounding medium was proposed decades ago by \cite{chugai93,chugai94,cdd95}.

 Many young SNe have shown an increase in X-ray emission consistent with the shock impacting a high-density medium a few years to decades after outburst. The X-ray emission in SN 1987A started rising after about 3 years. It was attributed to interaction with a dense HII region inside the shell \cite{cd95}. A hydrodynamic simulation that incorporates the effect of  the photoionizing photons from the progenitor star reveals that such a region is expected to form in SN 1987A (region (3) in Figure~\ref{fig:87afinal}).

 In SN 1996cr, the rise in X-ray emission began at an earlier time \cite{ddb10}, and in SN 2014C the shock-shell interaction may have started as early as half a year after outburst \cite{marguttietal17}, if not earlier \cite{qianetal25}. These all point to the existence of a high-density medium close to the star, and wind properties that change hundreds to thousands of years before stellar death, depending on the wind velocities. A rebrightening in X-rays is also noticeable to a lesser extent in SNe such as SN 2004dk \cite{pooleyetal19} (Type Ib) and SN 2011ja \cite{sayanetal13} (Type IIP).

In SN 1996cr, the presence of a dense surrounding medium was deduced from the rising X-ray and radio data \cite{baueretal08}. Relevant wind and CSM parameters were inferred by carrying out spherically symmetric numerical simulations, computing synthetic X-ray spectra, and comparing to observations \cite{ddb10}.

Simultaneously,  a high-resolution 484 ks Chandra grating observation of SN 1996cr demonstrated the presence of an asymmetry \cite{qbd19} in the structure of various lines, especially those of H-like Si and Fe (Figure 3 in \cite{qbd19}). This suggested an asymmetry in the SN ejecta, or the ambient medium, or both (Figure 15 in \cite{qbd19}).

\begin{figure}[H]
\includegraphics[width=\columnwidth]{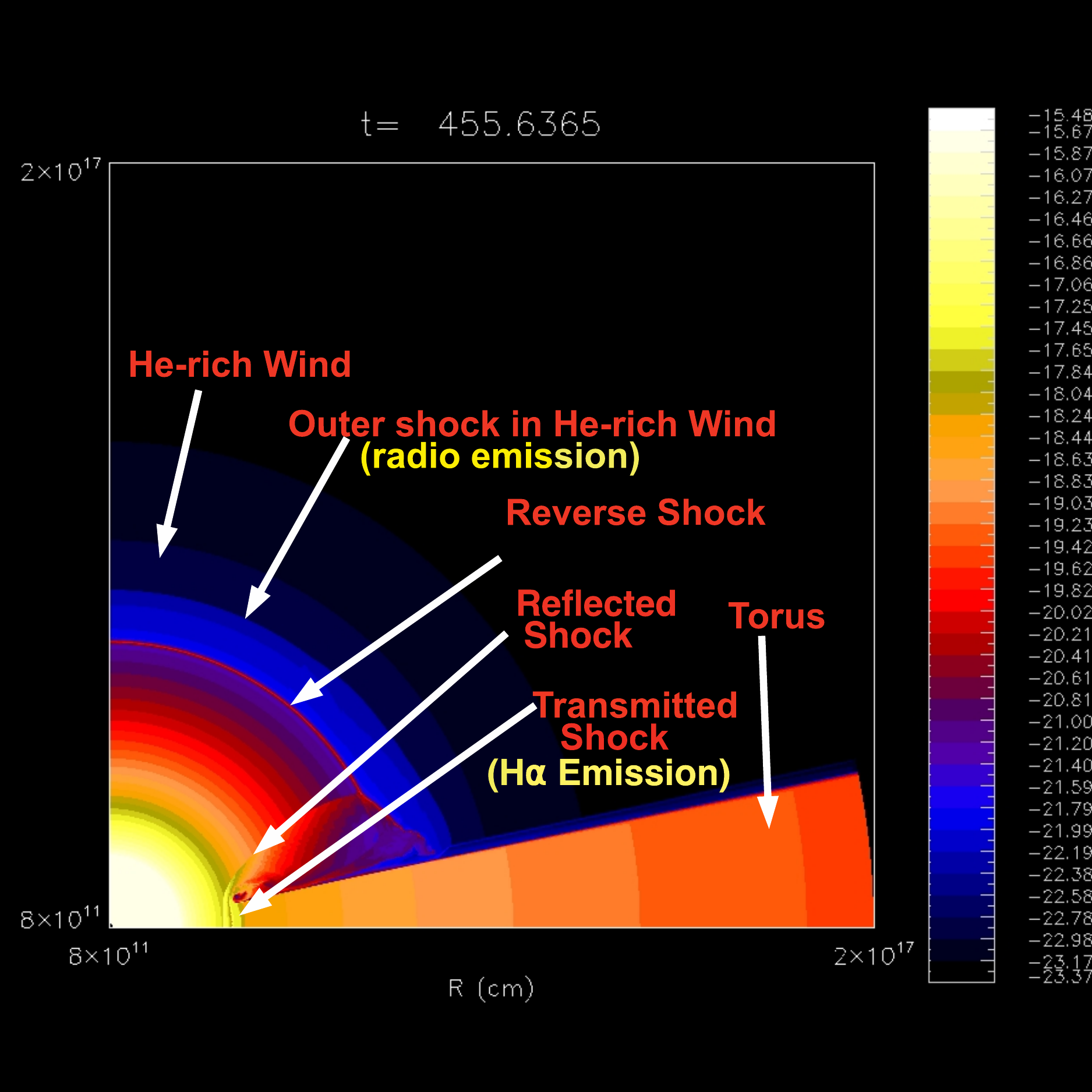}
\caption{The interaction of a SN shock wave with an equatorial torus, as may have happened in SN 2014c. The plot shows the density structure from a simulation carried out by the author to study the shock interaction with the equatorial torus in SN 2014c. The VH-1 numerical hydrodynamics code was used \cite{dcb96,db98,Dwarkadas2005,Dwarkadas2007c,ddb10}. The impact with the torus drives a transmitted shock into the torus, and a reflected shock back into the ejecta. The shock above and below the torus expands in a lower-density He-rich wind, and therefore, has a much higher velocity. The radio emission arises from this shock. The H$\alpha$ emission arises from the transmitted shock in the torus. The reflected shock and reverse shock in the He-rich wind tend to converge at a point. 
\label{fig:sntorus}}
\end{figure}

SN 2014c also showed an increase in the X-ray emission with time. The availability of data over the entire wavelength range, from radio to optical to X-rays, was essential to the interpretation of the emission. Rising X-ray emission was seen, first with Chandra and then with a combination of Chandra and NuStar telescopes\cite{marguttietal17}.  Ref. \cite{thomasetal22} found that while a high-density region, as indicated by the rising X-rays, was indeed required, having a spherical dense shell formed by a wind bubble did not fit the data. Instead, the velocities indicated by the radio and H$\alpha$ emission were very different, with VLBI data showing the radio shell expanding at $> 10,000$ km s$^{-1}$ \cite{bietenholzetal18,bietenholzetal21}, while the H$\alpha$ line width was around 2000 km s$^{-1}$ \cite{thomasetal22}. This indicated a two-component ambient medium, which \cite{thomasetal22} took to be an equatorial torus surrounded by lower-density wind material (Figure 11 in \cite{thomasetal22}). The collision of the SN shock wave with the dense torus leads to a slower transmitted shock within the torus, while the reflected shock expands back into the ejecta. Simultaneously, the SN shock above and below the torus, which is expanding in a lower-density He-rich wind, expands with a much higher velocity, giving rise to the radio emission (Figure \ref{fig:sntorus}). \cite{thomasetal22} postulated that such a medium could be formed by common envelope evolution from a binary progenitor. A few months later, \cite{brethaueretal22} formulated a similar model for SN 2014C.

\noindent

The formation of an equatorial disk around the SN 2001em was similarly deduced mainly from radio and X-ray data, along with published H$\alpha$ spectra \cite{chandraetal20}.

\section{Combining with Multiwavelength Data}
\label{sec:multiwv}

\begin{figure}[H]
\includegraphics[width=\textwidth]{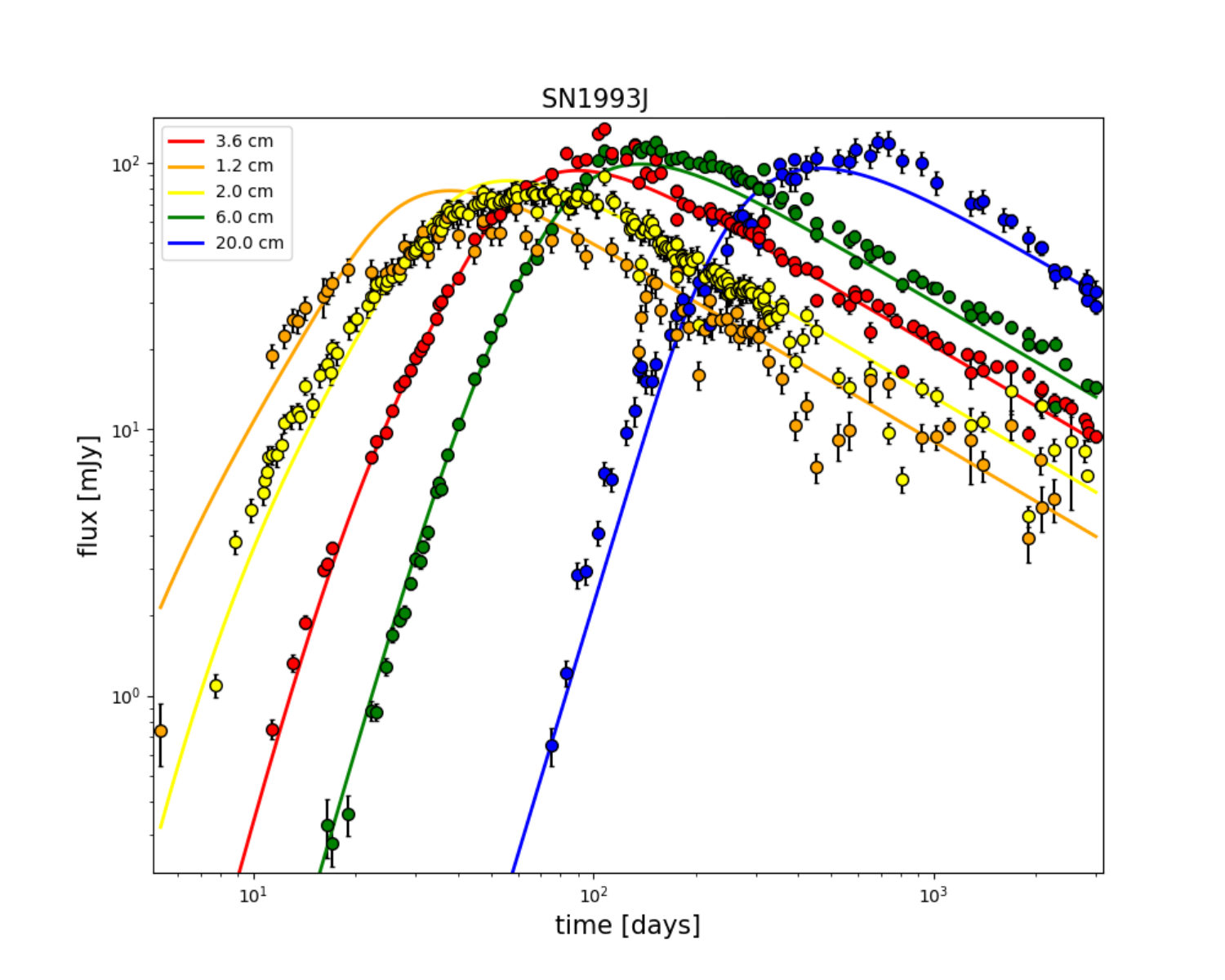}
\caption{Fits to the radio light curves of SN 1993J at various
  frequencies, using equations of \cite{chevalier82b} as
  modified by \cite{Weiler2002}, and an MCMC code developed by the author and his students.  
\label{fig:93jfit}}
\end{figure}

{\bf Radio:}  Radio light curves of SNe can also supply significant information about their progenitors, especially when combined with the X-ray data. A Markov-Chain Monte-Carlo (MCMC) code can be used to fit radio light curves and obtain fit parameters (Figure \ref{fig:93jfit}), as described in \cite{chevalier82b,Weiler2002}. Such fits have been carried out for many SNe  (see \cite{Weiler2002, chandraetal09b,chandraetal12b,drrb16,baerwayetal25, nayanaetal25}). 

Well-sampled radio light curves of young SNe can provide information on
various properties of the SN shock wave. The density slope of the medium in which the SN is
evolving can be obtained by
fitting the radio light curves at various frequencies (see
\cite{chevalier82b, Weiler2002}).  

The mass-loss rate $\dot{M}$ in $\msun$/yr$^{-1}$ can be derived from the turn-on of the radio absorption using the equation:

\begin{equation}
\frac{\dot{M}}{{v_w}_{1}} = 3. \times 10^{-6}\;({\tau}_{eff}^{0.5})\,m^{-1.5}\;{{v_i}_4}^{1.5}\,{{t_i}_{45}}^{1.5}\,\left[\frac{t}{t_i}\right]^{1.5m}\,T_4^{0.68}
\end{equation}

\noindent
where ${\tau}_{eff}$ is the effective optical depth \cite{Weiler2002}, $v_i$ is the initial velocity of the forward shock measured at time $t_i$, and $T_4$ is the temperature of the external medium in units of 10$^4$ K. \cite{Weiler2002} assumed that $t_i$ would be 45 days, but nowadays measurements are obtained much earlier during the evolution.

One can also estimate the circumstellar wind mass-loss rate from the radio emission \cite{Weiler2002} for Type Ib/c SNe using the following:

\begin{equation}
\frac{\dot{M}}{{v_w}_{1}} = 8.6 \times 10^{-9}\,\left[\frac{L_{6\,cm\,peak}}{10^{26}\; \mathrm{erg\,cm^{-1}\,Hz^{-1}}}\right]^{0.71}\;(t_{6\,cm\,peak} - t_0)^{1.14}
\end{equation}

For Type II SNe, a similar equation reads

\begin{equation}
\frac{\dot{M}}{{v_w}_{1}} = 1. \times 10^{-6}\,\left[\frac{L_{6\,cm\,peak}}{10^{26}\; \mathrm{erg\,cm^{-1}\,Hz^{-1}}}\right]^{0.54}\;(t_{6\,cm\,peak} - t_0)^{0.38}
\end{equation}

 \noindent
 where the time from an initial explosion time $t_0$ is measured in days. \cite{Weiler2002} assumed a wind velocity of 10 km s$^{-1}$ for Type Ib/c SNe, although it is now accepted that they arise from Wolf-Rayet stars with a wind velocity of order 1000--3000 km s$^{-1}$. It is assumed that the opacity at the peak of the 6 cm emission ${\tau}_{6cm} \approx 1$.
 
  A combination of radio and X-ray light curves can provide
information on the density and, if used carefully, the ionization
structure of the ambient medium \cite{meunieretal13}. Deceleration of the SN shock wave, and asymmetry in the medium, can be inferred by a careful examination of radio images and light curves even before the structure can be resolved, as in SN 1987A \cite{gaensleretal97}.

A combination of optical, X-ray, radio and IR light curves
provides information on density inhomogeneities, presence of dust
shells, heating mechanisms, and evidence of
outbursts. A few IIns suggest enhanced mass-loss
prior to explosion \cite{dg12,foxetal13, drrb16, szalaietal21,dwarkadas25} using multi-wavelength data. 

Often, the mass-loss rate is derived from the optical/IR/UV emission, assuming that the luminosity is some fraction $\epsilon$ of the total available luminosity at that epoch \cite{salamancaetal98,smithetal08,scsff10,milleretal10b,andrewsetal17, bilinskietal18, kumaretal19, bilinskietal20,sa20, dickinsonetal24}:

\begin{equation}
L_{opt/IR/UV} = 0.5\,\epsilon \; \frac{\dot{M}}{v_w}\;v_{sh}^3
\end{equation}

This makes several assumptions. It assumes that the SN is evolving in a wind whose density decreases as r$^{-2}$. As evidenced from Figure \ref{fig:lcmdot}, this is not necessarily true. The coefficient $\epsilon$ is clearly unknown, although values between 0.2 and 1 are commonly used. It is unlikely that the shock velocity is a constant over a large time period. In the Chevalier self-similar model mentioned earlier, it varies as $t^{m-1}$. If $m \approx 0.86$ then  the velocity varies as $t^{-0.14}$. The wind velocity $v_w$ is not always known, although it is often taken to be the width of narrow lines seen in the optical spectrum, particularly in Type IIn SNe. Also, the luminosity employed may not be the total bolometric luminosity. Using a luminosity in a limited waveband will lead to an artificially low mass-loss rate. For these reasons, obtaining a mass-loss rate from the X-ray light curve and spectrum is preferable, as shown earlier.

\section{Tying Wind Parameters to Progenitors} 
\label{sec:windpar}
As a first assumption, this is conducted assuming spherical symmetry, although asymmetries  will often be revealed in the process. The wind parameters for different types of stars are somewhat well calibrated \cite{pvn08,mj11,smith14}, allowing the mass loss rate to be tied to the progenitor \cite{vanloon25}. Mass loss rates for massive stars are a function of stellar parameters \cite{vinketal01,vink2008,vs21} such as radius and luminosity, and can often be related to them. In complicated situations, appropriate choices need to be made. Expansion in a  wind bubble can enable wind parameters to be obtained \cite{ddb10} by combining judicious choices with available data. While every situation is different, reasonable estimates of the mass loss rates and the progenitor are often possible \cite{ddb10,chandraetal12b, sayanetal13, maedaetal14,drrb16, sayanetal16,qbd19,chandraetal20,thomasetal22, gd26}, especially when combined with radio data.

Investigations of X-ray lines in high-resolution spectra, if available, can reveal the presence of asymmetries in the SN explosion or surrounding medium, as determined for SN 1996cr \cite{qbd19}. Anomalies in the mass loss rate or density distribution may often suggest binary evolution \cite{om17}.

%%%%%%%%%%%%%%%%%%%%%%%%%%%%%%%%%%%%%%%%%%
\section{Discussion}
\label{sec:disc}

As shown in this paper, X-ray emission from a SN can be used to constrain many properties of the progenitor. This includes the mass-loss rate of the progenitor star before explosion. The fact that the SN shock wave can evolve up to 1000 times faster than the pre-SN wind enables one to follow, in real time, the evolution of the wind mass-loss rate over decades and centuries. The mass-loss rate can be derived by various means as outlined in the text, and it is essential to derive it by multiple means to verify its accuracy. It should be emphasized that the mass-loss rate generally occurs in equations in combination with the velocity, i.e., as the parameter $\dot{M}/v_w$. Therefore, while we emphasize the mass-loss, it is the ratio of mass-loss rate to wind velocity that is obtained, and obtaining the mass-loss does assume that the velocity is known. It is also plausible that it is the mass-loss rate that is constant, while the velocity is varying.  The X-ray emission can also be used to constrain the abundance of various elements in the progenitor, although the abundance that is uncovered may depend on the fraction of the ejecta that is shocked. Finally, the X-ray emission, combined with other available multi-wavelength emission, can be used to constrain the structure and density profile of the surrounding medium, as in SN 1987A \cite{cd95}, SN 1996cr \cite{ddb10,qbd19}, and SN 2014C \cite{thomasetal22, brethaueretal22}.

Thermal X-rays provide  the most direct signature of the circumstellar density, since the emission is directly proportional to the density. They also provide the most accurate measure of the density, as described herein. While the radio emission from SNe is also used to compute the density, the emission mechanism is synchrotron emission, due to electrons spiraling around a magnetic field. The relationship to the circumstellar density is indirect, as it is for non-thermal X-ray emission. Optical emission, such as in the H$\alpha$ line, is also proportional to the density squared. However, the mass of material required to produce the line does not need to be high, and could be a fraction of a solar mass. Thus, optical emission may arise from smaller high-density clumps rather than the bulk of the material, and the properties derived may not reflect the bulk of the material.

%%%%%%%%%%%%%%%%%%%%%%%%%%%%%%%%%%%%%%%%%%
%\section{Conclusions}

%%%%%%%%%%%%%%%%%%%%%%%%%%%%%%%%%%%%%%%%%%
\vspace{6pt} 

%%%%%%%%%%%%%%%%%%%%%%%%%%%%%%%%%%%%%%%%%%

%%%%%%%%%%%%%%%%%%%%%%%%%%%%%%%%%%%%%%%%%%
%\authorcontributions{N/A}

\funding{This research received no external funding.}

\institutionalreview{Not applicable.}

\dataavailability{All data used are publicly available or provided in the paper.}

%\acknowledgments{}

\conflictsofinterest{The author declares no conflicts of interest.} 

%%%%%%%%%%%%%%%%%%%%%%%%%%%%%%%%%%%%%%%%%%
%% Optional

%% Only for journal Encyclopedia
%\entrylink{The Link to this entry published on the encyclopedia platform.}

%%%%%%%%%%%%%%%%%%%%%%%%%%%%%%%%%%%%%%%%%%

% Please provide the correct journal abbreviation (e.g., according to the “List of Title Word Abbreviations” http://www.issn.org/services/online-services/access-to-the-ltwa/).
% Citations and References in Supplementary files are permitted provided that they also appear in the reference list here. 

%=====================================
% References, variant A: external bibliography
%=====================================
%\bibliography{apj-jour,paper1,paper2}

\begin{adjustwidth}{-\extralength}{0cm}
\reftitle{References}

%=====================================
% References, variant B: internal bibliography
%=====================================

% If authors have biography, please use the format below
%\section*{Short Biography of Authors}
%\bio
%{\raisebox{-0.35cm}{\includegraphics[width=3.5cm,height=5.3cm,clip,keepaspectratio]{Definitions/author1.pdf}}}
%{\textbf{Firstname Lastname} Biography of first author}
%
%\bio
%{\raisebox{-0.35cm}{\includegraphics[width=3.5cm,height=5.3cm,clip,keepaspectratio]{Definitions/author2.jpg}}}
%{\textbf{Firstname Lastname} Biography of second author}

% For the MDPI journals use author-date citation, please follow the formatting guidelines on http://www.mdpi.com/authors/references
% To cite two works by the same author: \citeauthor{ref-journal-1a} (\citeyear{ref-journal-1a}, \citeyear{ref-journal-1b}). This produces: Whittaker (1967, 1975)
% To cite two works by the same author with specific pages: \citeauthor{ref-journal-3a} (\citeyear{ref-journal-3a}, p. 328; \citeyear{ref-journal-3b}, p.475). This produces: Wong (1999, p. 328; 2000, p. 475)

%%%%%%%%%%%%%%%%%%%%%%%%%%%%%%%%%%%%%%%%%%
%% for journal Sci
%\reviewreports{\\
%Reviewer 1 comments and authors’ response\\
%Reviewer 2 comments and authors’ response\\
%Reviewer 3 comments and authors’ response
%}
%%%%%%%%%%%%%%%%%%%%%%%%%%%%%%%%%%%%%%%%%%
\PublishersNote{}
\end{adjustwidth}
%\isPreprints{}{% This command is only used for ``preprints''.
%\end{adjustwidth}
%} % If the paper is ``preprints'', please uncomment this parenthesis.
\end{document}